\documentclass[aps,prl,twocolumn,superscriptaddress,nofootinbib]{revtex4-2}
\usepackage{amsmath, bm}

\newcommand{\ce}[1]{Eq.~(\ref{#1})}

\newcommand{\ct}[1]{{Table~\ref{#1}}}

\DeclareMathAlphabet{\pazocal}{OMS}{zplm}{m}{n}

\newcommand{\oo}{\ensuremath{\pazocal{O}}}

\usepackage{color}
\usepackage[normalem]{ulem}
\usepackage{cancel}
\usepackage{txfonts}

\newcommand{\pT}{\ensuremath{P_T}\xspace}
\newcommand{\jpsi}{\ensuremath{J/\psi}\xspace}
\newcommand{\psip}{\ensuremath{\psi(2S\!)}\xspace}
\newcommand{\chic}{\ensuremath{\chi_c}\xspace}
\newcommand{\chico}{\ensuremath{\chi_{c1}}\xspace}
\newcommand{\chict}{\ensuremath{\chi_{c2}}\xspace}

\usepackage{graphicx}
\usepackage{xspace}
\usepackage{array}
\usepackage{float} %
\usepackage{subfigure}
\usepackage{comment}
\usepackage{booktabs}
\usepackage{multirow}
\usepackage{adjustbox}
\usepackage{rotating}
\usepackage{lineno}
\usepackage{enumerate}
\usepackage{enumitem}
\setlist[enumerate]{noitemsep, topsep=0pt, parsep=0pt, partopsep=0pt,leftmargin=1em}

\usepackage{ifpdf}
\ifpdf
\usepackage[pdftex]{hyperref}
\else
\usepackage[hypertex]{hyperref}
\fi

\hypersetup{
   pdftitle={},%
   pdfauthor={},%
   pdfsubject={},%
   pdfkeywords={},%
   pdfstartview={},%
   bookmarksopen=true, breaklinks=true, debug=true, %
   colorlinks=true, linkcolor=red, citecolor=blue, urlcolor=blue
}

\begin{document}
\title{A consistent \texorpdfstring{$\oo (v^2)$}{O(v2)} NRQCD description of decays and high-\texorpdfstring{$\pT$}{pT} prompt
production
of charmonia}

\author{Valerio Bertone}
\affiliation{Universit\'e Paris-Saclay, CEA, IRFU, 91191 Gif-sur-Yvette, France}

\author{Jean-Philippe Lansberg}
\affiliation{Universit\'e Paris-Saclay, CNRS, IJCLab, 91405 Orsay, France}

\author{Kate Lynch}
\affiliation{Universit\'e Paris-Saclay, CNRS, IJCLab, 91405 Orsay, France}
\affiliation{School of Physics, University College Dublin, Dublin 4, Ireland}

\author{Tanjona R. Rabemananjara}
\affiliation{Universit\'e Paris-Saclay, CNRS, IJCLab, 91405 Orsay, France}


\begin{abstract}
{We present a global determination of the six long-distance matrix elements (LDMEs) within non-relativistic QCD
(NRQCD) appearing in the production of prompt $\jpsi$ up to $\oo (v^2)$ using high-$P_T$ LHC data for direct and
feed-down contributions from $\chi_c$ and $\psi(2S)$. Using a leading-power fragmentation framework at NLO+NLL,
we fit $\psi(2S)/J/\psi$ yield ratios from ATLAS and CMS and $\psi(2S)$ CMS polarization data at $\sqrt{s} = 13$
TeV, injecting one decay width per charmonium.
The fit LDMEs provide excellent \emph{predictions} of the CMS and ATLAS cross-section data up to $P_T=360$ GeV
for the $J/\psi$, as well as the $J/\psi$ polarization data from CMS.
The obtained excellent agreement with production and decay data demonstrates that $\oo (v^2)$ NRQCD is sufficient
to describe $J/\psi$, $\psi(2S)$, and $\chi_c$ hadroproduction at large $P_T$.
}
\end{abstract}

\maketitle

\paragraph{Introduction.}

In~\cite{Bertone:2025jex} we demonstrated that, for prompt $\psip$
production at the LHC, leading-power (LP) predictions at next-to-leading order (NLO) in perturbative QCD, supplemented by next-to-leading
logarithmic (NLL) resummation of collinear logarithms, along with $\oo (v^2)$ relativistic corrections are able to simultaneously describe the $\pT$-differential
cross sections measured by ATLAS~\cite{ATLAS:2023qnh} and CMS~\cite{CMS:2017dju},
\textit{and} the polar anisotropy parameter $\lambda_\theta$ measured by
CMS~\cite{CMS:2024igk}, without the need of $\oo(v^{n>2})$ contributions.
The very good agreement with $\psip$ LHC data comes from charm fragmentation, found to  account for 30\%-50\% of the yield, and from the gluon fragmentation, which receives very large relativistic corrections. When combined, these contributions produce a slightly
transversely polarized $\psip$ yield in agreement with CMS data.

In this work, we focus on the more complex case of $\jpsi$ production, whose measurements~\cite{ATLAS:2023qnh} extend to
$P_T=360$~GeV, but include significant
feed down (FD) from $\chic$ and $\psip$.
{Our strategy is to get a handle on:} 
\begin{enumerate}
\item the $\chic$ FD from  CMS and ATLAS $\psip / \jpsi$ yield ratios
measured up to $P_T=120$~GeV --- whereas direct $\chi_c$ measurements~\cite{ATLAS:2014ala} barely reach $P_T=25$~GeV, where
next-to-LP corrections can be sizable~\cite{Ma:2014svb},

\item the relative contributions of charm and gluon channels from
the CMS $\psip$ polarization data also up to $P_T=120$~GeV,

\item the normalizations of the charmonium cross sections from
one decay width per charmonium.
\end{enumerate}
Cross sections are therefore used in our fit only through ratios.

\paragraph{Methodology.}
Up to $\oo (v^2)$, {direct} $\jpsi$ and $\psip$ (collectively denoted as $\psi$) cross sections only differ by the value of the LDMEs. For $\chic$,
the {fragmentation functions (FFs)} are different and, in fact, one of them involves a color-octet (CO) state. The LP $P_T$-differential cross section
for prompt {$\jpsi$} production in proton-proton collisions at $\oo (v^2)$ in NRQCD is obtained {from} 
\begin{align}\label{eq:LP}
   \frac{d\sigma^{(\psi, \chic)}}{dP_{T}^{\jpsi}} = \sum_{i,j,k}^{m,n} &f_{i}(\mu_f) \otimes f_{j}(\mu_f)
   \otimes \frac{d\hat{\sigma}_{i j \to k X }(\mu_f,\mu_F)}{dP_{T}} \otimes d_{k/m}^{n} (\mu_F) \nonumber \\& 
   \times
   \left\langle \oo_n^{(\psi, \chic)} \left(m\right) \right\rangle \mathcal{B}^{(\psi, \chic)}_{\jpsi}
   \big\lvert_{\pT^{\jpsi}=\frac{M_{\jpsi}}{M_{(\psi, \chic)}} \pT^{(\psi, \chic)}}\,,
\end{align}
where
$f$ are the collinear Parton Distribution Functions (PDFs), $d\hat{\sigma}$ are the partonic
cross sections {in the zero-mass variable-flavor number scheme (ZM-VFNS)}, and $d^n$ are the FFs amputated from the LDMEs.
{For $\psi$, we consider $m= {}^{3}S_1^{[1]}$ and  $n\in \{0,2\}$ and, for $\chi_{cJ}$, $ m\in \{{}^{3}P_J^{[1]},{}^{3}S_1^{[8]} \}$ and $n=0$.}
 Wherever relevant for FD contributions,
$\mathcal{B}$ denotes the branching fraction and  we account for the $P_T$ shift induced by the decay as indicated.

In \ce{eq:LP}, PDFs and FFs are evaluated at the scales
$\mu_f$ and $\mu_F$, respectively.
NRQCD provides factorized expressions for the FFs at their initial scales $\mu_0$ in terms of the
short-distance coefficients $d^n$ and the long-distance matrix elements $\langle \oo_n(m) \rangle$.
{Like in~\cite{Bertone:2025jex}, we
use all existing knowledge on FF initial conditions~\cite{Braaten:1993mp,Braaten:1995cj,Bodwin:2014bia,Ma:2013yla,Ma:2015yka,Zhang:2017xoj,Zheng:2019dfk,Feng:2021qjm} for $\psi$ up to $\oo(\alpha_s^3 v^2)$ for gluon and charm channels. All the initial conditions that we employ are indeed known at NLO in $\alpha_s$, except for the $g \to\, ^3S_1^{[1]}gg$ FF and the polarized $c \to\, ^3S_1^{[1]}c$ FF. For $\chi_c$, we work at $\oo(\alpha_s^2 v^2)$, i.e., NLO for the $g \to\, ^3S_1^{[8]}$ FF~\cite{Ma:2013yla,Ma:2015yka}. } %
We evolve them {up to $\mu_F$  using the DGLAP equations  at NLL accuracy
after a Mellin transform  using \texttt{MELA}~\cite{Bertone:2015cwa} 
or {directly using} \texttt{APFEL++}~\cite{Bertone:2017gds,Bertone:2013vaa}. Light-flavor FFs are generated by evolution. We then}
fold the evolved FFs with $d\hat{\sigma}$ computed up to $\oo (\alpha_s^3)$ (NLO) using INCNLO~\cite{Werlen:INCNLO}.
{Doing so, we achieve  NLO+NLL accuracy.}
%

%
\begin{table*}[t]
\centering
\renewcommand{\arraystretch}{1.35}
\setlength{\tabcolsep}{4pt}
\caption{
    Best-fit values of the 6 LDMEs up to $\oo(v^2)$, the 3 derived leptonic decay widths,
    and the 2 nuisance parameters, with asymmetric $1\sigma$ uncertainties from profiling
    $\chi^2_{\rm tot}$. The goodness of fit is quoted as $\chi^2_{\rm data}/ n_{\rm data}$.
}
\label{tab:best_fit_parameters}
\vspace{10pt}
\begin{tabular*}{\textwidth}{@{\extracolsep{\fill}}cccccccccccc@{}}
\toprule
$\langle v^2_{\psip}\rangle$ &
$\langle v^2_{\jpsi}\rangle$ &
$|R_{\psip}(0)|^2$ &
$|R_{\jpsi}(0)|^2$ &
$|R^\prime_{\chi_{c}} (0)|^2$ &
$\left\langle \oo_{0}^{\chi_{c 0}}\left({ }^3 S_1^{[8]}\right)\right\rangle$ &
$\Gamma^{ee}_{\psip}$ &
$\Gamma^{ee}_{\jpsi}$ &
$\Gamma^{\gamma \gamma}_{\chict}$ &
$\delta_s$ &
$\delta_s^\prime$ &
$\chi^2_{\rm data}/n_{\rm data}$ \\
   & 
   & 
   [GeV$^3$] & 
   [GeV$^3$] & 
   [GeV$^5$] & 
   [$\times 10^{-2}$ GeV$^3$] &
   [keV] & 
   [keV] & 
   [keV] &
   & 
   & 
   \\
\midrule
   $0.38^{+0.06}_{-0.06}$ & 
   $0.22^{+0.07}_{-0.03}$ & 
   $0.93^{+0.17}_{-0.14}$ & 
   $1.36^{+0.21}_{-0.18}$ & 
   $0.08^{+0.01}_{-0.01}$ & 
   $0.36^{+0.05}_{-0.05}$ &
   $2.33^{+0.06}_{-0.06}$ & 
   $5.53^{+0.14}_{-0.14}$ & 
   $0.57^{+0.09}_{-0.09}$ &
   $0.07^{+0.16}_{-0.16}$ & 
   $0.05^{+0.06}_{-0.06}$ & 
   $0.75$ \\
\bottomrule
\end{tabular*}
\end{table*}
\begin{figure*}[tb]
    \centering 
   \includegraphics[width=0.32\linewidth]{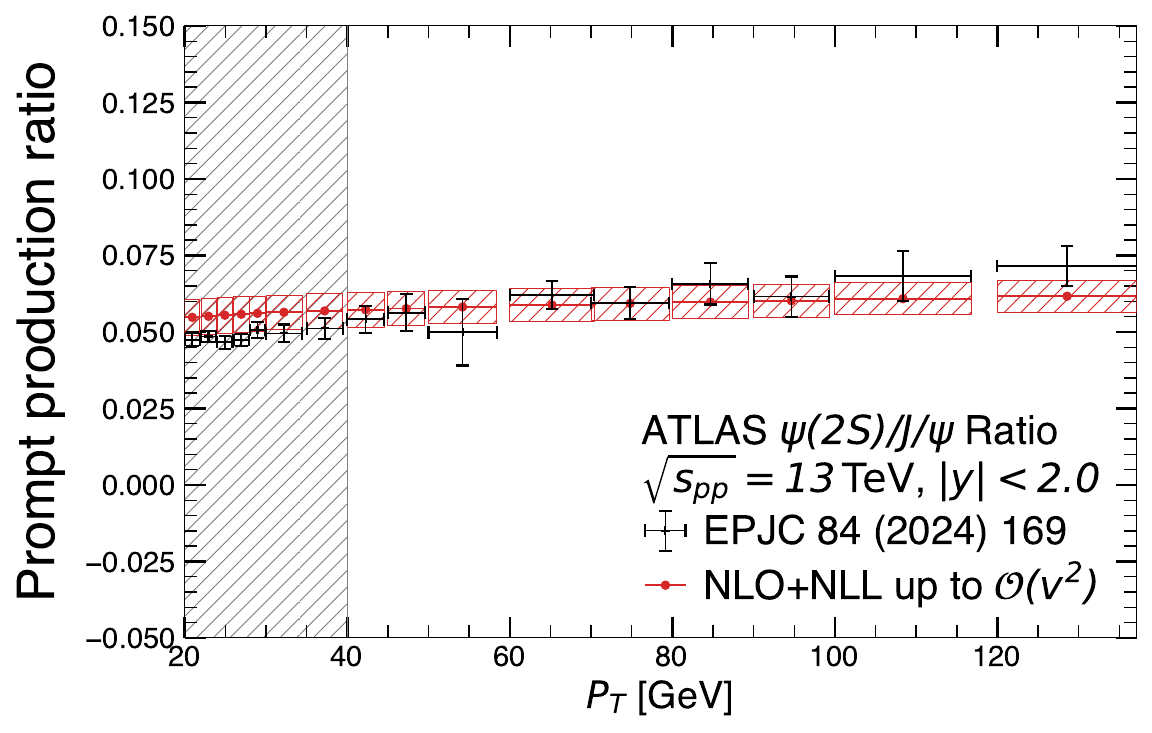}
   \llap{(a)\hspace{5.4cm}}\includegraphics[width=0.32\linewidth]{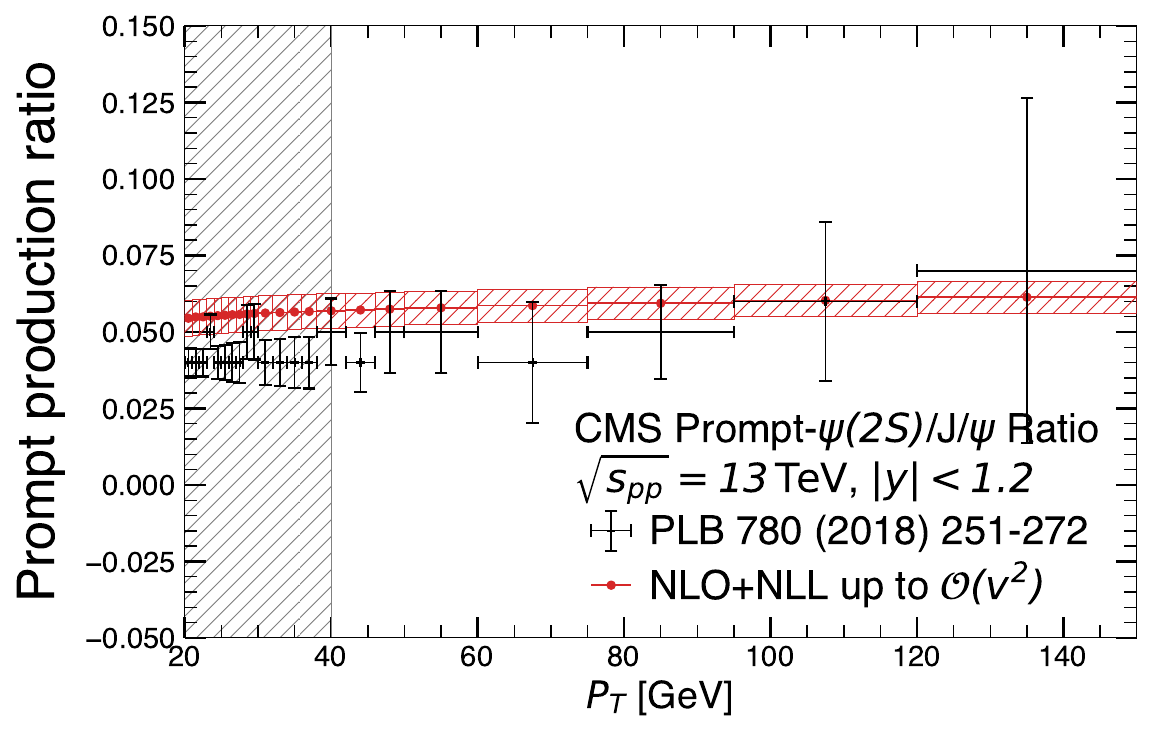}
   \llap{(b)\hspace{5.4cm}}\includegraphics[width=0.32\linewidth]{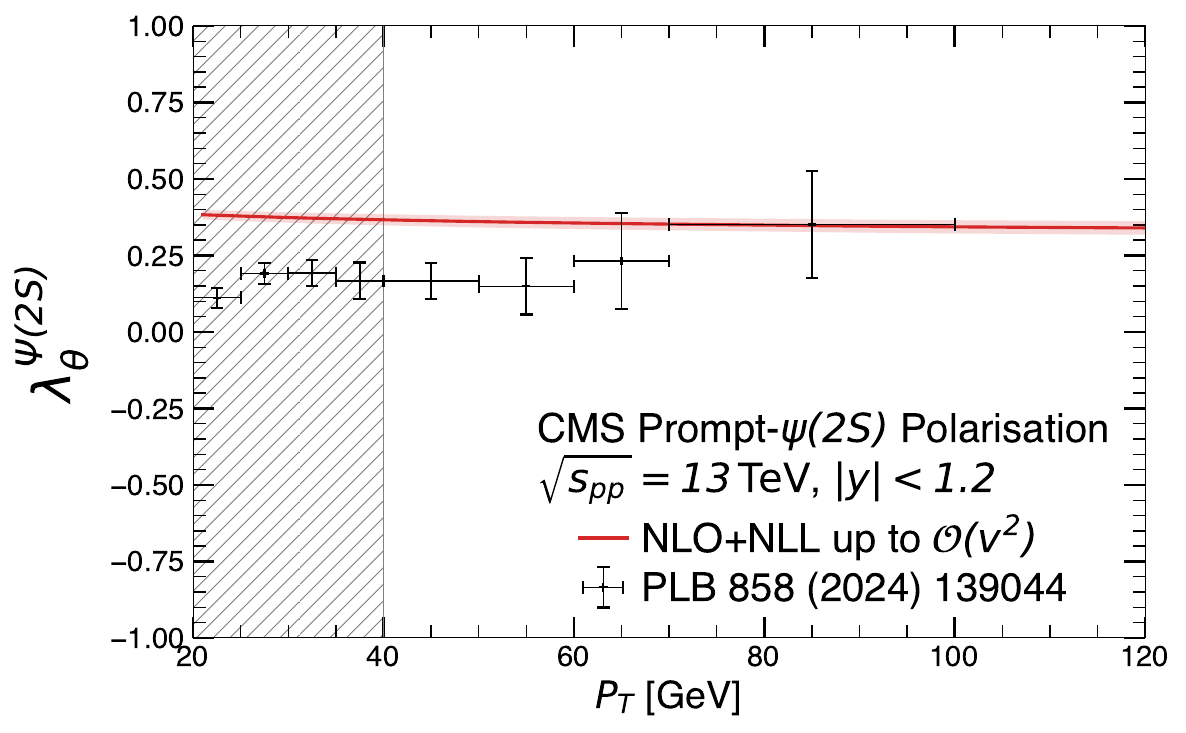}
   \llap{(c)\hspace{5.4cm}}
    \caption{{Comparisons of  NLO+NLL NRQCD up to $\oo(v^2)$, using the values of \ct{tab:best_fit_parameters}, to the 3 production data sets used in our fit: (a) ATLAS~\cite{ATLAS:2023qnh} and (b) CMS~\cite{CMS:2017dju} prompt $\psip/\jpsi$ yield ratio  and  (c)  CMS prompt-$\psip$ polar anisotropy $\lambda^{\psip}_\theta$~\cite{CMS:2024igk}. The red band shows the $1\sigma$
    envelope of the fit, obtained by propagating the fit-parameter covariance matrix through the theoretical
    expressions. The zone in gray where NLP contributions can be important is excluded from the fit.}}
    \label{fig:fit_results}
\end{figure*}

As mentioned, we fit (i) the high-$\pT$ $y$-integrated $\psip/\jpsi$ yield-ratio~\cite{ATLAS:2023qnh,CMS:2017dju}, (ii) $\lambda_\theta^{\psip}$
in the helicity frame~\cite{CMS:2024igk}, both satisfying  $\pT \geq 40$~GeV, and (iii) the decay width  values for $\Gamma^{ee}_\psi$ and $\Gamma^{\gamma \gamma}_{\chict}$ {from the PDG}~\cite{ParticleDataGroup:2024cfk}. This amounts to $n_{\rm data} = 24$ experimental data points, $y_i^{\rm data}$, whose statistical and systematic uncertainties are added in quadrature as Gaussian and uncorrelated {into $\Delta_i$}.

By virtue of the vacuum-saturation approximation and heavy-quark-spin symmetry, {respectively} valid up to $\oo (v^{4,2})$, the LDMEs entering Eq.~\eqref{eq:LP}
satisfy the following relations~\cite{Bodwin:1994jh}: $\left\langle \oo^{\psi}_0\left({ }^3 S_1^{[1]}\right)\right\rangle = \frac{3 N_c}{2 \pi} |R_{\psi} (0)|^2$, $\left\langle \oo_{0}^{\chi_{c J}}\left({ }^3 P_J^{[1]}\right)\right\rangle  =(2 J+1) \frac{3 N_c}{2 \pi} |R^{\prime}_{\chic}(0)|^2$, and $\left\langle \oo_{0}^{\chi_{c J}}\left({ }^3 S_1^{[8]}\right)\right\rangle =(2 J+1)\left\langle \oo_{0}^{\chi_{c 0}}\left({ }^3 S_1^{[8]}\right)\right\rangle$,
where $|R_{\psi}(0)|^2$ is the {radial} wavefunction at the origin and $|R_{\chi_c}^{\prime}(0)|^2$ its derivative. 
Using the Gremm-Kapustin relation~\cite{Gremm:1997dq}, the $\psi$ $\oo( v^2)$ LDMEs
 can be expressed in terms of the average velocity squared as
$\left\langle \oo_2^{\psi} \left({ }^3 S_1^{[1]}\right) \right\rangle = m_c^2 \langle v_\psi^2 \rangle \left\langle \oo_0^{\psi} \left({ }^3 S_1^{[1]}\right) \right\rangle$. As such, our fit of the 6  LDMEs up to $\oo (v^2)$ is tantamount to a fit of
$|R_{\psi}(0)|^2$, $|R^{\prime}_{\chic}(0)|^2$,  $\langle v^2_{\psi} \rangle$, and one CO LDME $\left\langle \oo_0^{\chi_{c 0}}\left({ }^3 S_1^{[8]}\right)\right\rangle$.

Except for the $\chic$ contributions, the computations follow the same line as~\cite{Bertone:2025jex},
where we identified, through a detailed analysis of the theoretical uncertainties, that the largest is by
far that from the renormalization scale $\mu_{R,0}$  in the  gluon FF which is only known at LO~\cite{Braaten:1995cj,Zhang:2017xoj}.
As such, it deserves a specific treatment via a nuisance parameter $\delta_s${, which} allows us to interpolate
between lower and upper cross sections, $d \sigma^{\psi, \pm}$, via
\begin{equation}
    d \sigma^\psi (\delta_s) = d\sigma^{\psi, 0} + \delta_s \left[ \Theta(\delta_s) \cdot \Delta d\sigma^{\psi, +}
    + \Theta(-\delta_s) \cdot \Delta d\sigma^{\psi, -} \right]
    \label{eq:delta_s}
\end{equation}
where $\Delta d\sigma^{\psi, \pm} \equiv \pm (d\tilde{\sigma}^{\psi, \pm} - d\tilde{\sigma}^{\psi, 0})$, with $d\tilde{\sigma}^{\psi, 0}$ the central value.
This procedure {is necessary} to take into account the common perturbative pieces in the numerator and 
denominator of the ratios we fit.
As such, we use the same $\delta_s$ for the $\jpsi$ and $\psip$ cross sections. Strictly speaking, we should have associated the nuisance parameter only with the gluonic contribution. Doing it at
the cross-section level, like in~\ce{eq:delta_s}, introduces a  correction scaling like $\delta_s v^4$ which is negligible as our $\delta_s$ is found
to be very close to zero.
In general, $|\delta_s|$ below unity indicates good  {agreement between fit data and  theory within the theory uncertainties}. In particular, values close to zero -- as ours -- suggest that the central scale gives the best data
description.

To  constrain the normalizations, instead of using cross-section values, 
we use the decay widths $\Gamma^{ee}_{\psi}$ and
$\Gamma^{\gamma \gamma}_{\chict}$ up to $\oo (\alpha_s, v^2)$
\begin{align}
    \Gamma^{\gamma \gamma}_{\chict} & = \frac{64 \alpha^2}{45 m_c^4} |R^\prime_{\chi_{c}} (0)|^2
    \left(1 - \frac{16}{3} \frac{\alpha_s(\mu_{R})}{\pi} \right), \label{eq:cdecay} \\
    \Gamma^{ee}_{\psi} & = \frac{4\alpha^2 e_c^2}{M^2_{\psi}} |R_{\psi}(0)|^2 \left(1 - \frac{16}{3}
    \frac{\alpha_s(\mu_R)}{\pi} - \frac{1}{3}\langle v^2_{\psi} \rangle\right).
    \label{eq:qdecay}
\end{align}
The NLO decay widths also suffer from large scale uncertainties, which we treat similarly to
Eq.~(\ref{eq:delta_s}) with another nuisance parameter $\delta_s^\prime$.
In \ce{eq:qdecay}, we use the decay width $\Gamma^{ee}_{\psi}$ written in terms of $M_{\psi}$ (rather
than $m_c$), as proposed in~\cite{Bodwin:2006yd}. 

As for $\langle v^2_{\psi} \rangle$, we let them vary independently within the ranges
$\langle v^2_{\psip} \rangle \in [0.25, 0.5]$ and $\langle v^2_{\jpsi} \rangle \in [0.15, 0.35]$, subject to the ordering constraint
$\langle v^2_{\psip} \rangle > \langle v^2_{\jpsi} \rangle$, consistent with the Gremm--Kapustin
relation
~\cite{Gremm:1997dq}.
The cross sections appearing in the ratios and the polarizations are interpolated in $\langle v^2_{\psi} \rangle$, allowing for a continuous variation in the fit.

Our six optimal LDMEs are obtained by minimizing
\begin{equation}
    \chi^2_{\rm tot} = \chi^2_{\rm data} + \delta_s^2 +
    \delta^{\prime 2}_s,
    \label{eq:chi2}
\end{equation}
where $\chi^2_{\rm data} = \sum^{n_{\rm data}}_{i=1} [(y_i^{\rm data} - y_i^{\rm th})/\Delta_i]^2$ runs over the
$\psip/\jpsi$ yield ratio vs.~$\pT$, $\lambda_\theta^{\psip}$ vs.~$\pT$, $\Gamma^{ee}_{\jpsi}$, $\Gamma^{ee}_{\psip}$, and $\Gamma^{\gamma \gamma}_{\chict}$. 
The minimization is performed using a limited-memory BFGS bound-constrained algorithm~\cite{Byrd:1995:LBFGSB}, scanning 
several initial parameter values to avoid local minima.
Throughout this work, we use the CT18 PDFs~\cite{Hou:2019efy}. 
%

\begin{figure*}[tb]
    \centering
    \includegraphics[width=1.0\linewidth]{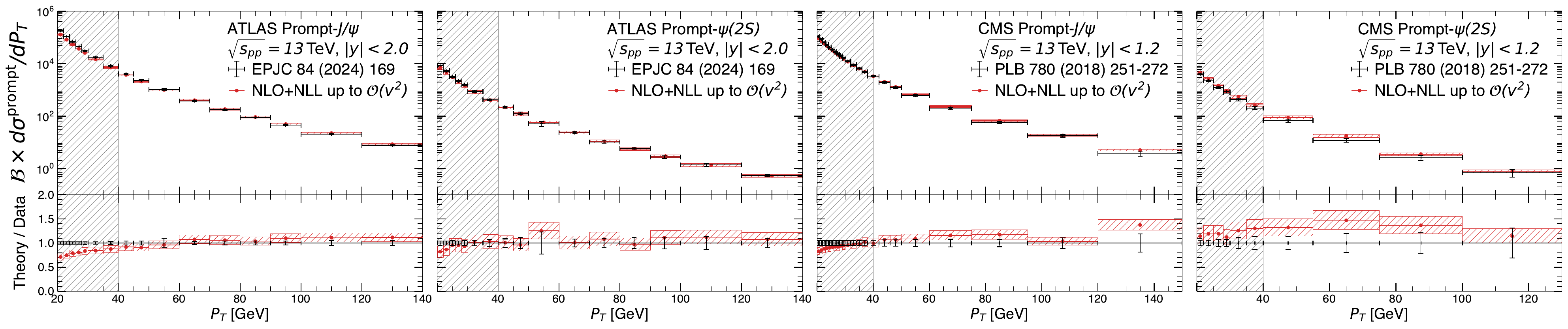}
    \caption{$\pT$-differential cross sections 
    at LP NLO+NLL up to $\oo(v^2)$ for prompt $\jpsi$ and $\psip$ production,     obtained with our six fit LDMEs (see Table~\ref{tab:best_fit_parameters}) compared to the ATLAS~\cite{ATLAS:2023qnh}
    and CMS~\cite{CMS:2017dju} measurements. Unlike the production ratio and the $\psip$ polarization
    shown in Fig.~\ref{fig:fit_results}, these absolute spectra are not included in the fit. The lower panels show
    the theory-to-data ratio. The zone in gray where NLP contributions can be important is just shown for illustration.
    }
    \label{fig:prediction_results}
\end{figure*}

\paragraph{$\oo (v^2)$ LDME fit results.}
Our fit results are collected in Table~\ref{tab:best_fit_parameters}. The minimization yields
$\chi^2_{\rm data}/n_{\rm data} = 0.75$, indicating that the LP NLO+NLL NRQCD calculations,
up to $\oo (v^2)$, provide a good description of the production data used in the fit, as shown
in Fig.~\ref{fig:fit_results}.
The same applies to the decay widths in Table~\ref{tab:best_fit_parameters}, which reproduce the PDG values~\cite{ParticleDataGroup:2024cfk}
of $\Gamma^{ee}_{\psip} = 2.33 \pm 0.04~\mathrm{keV}$, $\Gamma^{ee}_{\jpsi} = 5.53 \pm 0.10~\mathrm{keV}$, and
$\Gamma^{\gamma \gamma}_{\chict} = 0.57 \pm 0.04 ~\mathrm{keV}$.

We find $(\langle v^2_{\psip}\rangle, \langle v^2_{\jpsi}\rangle) = (0.38^{+0.06}_{-0.06},0.22^{+0.07}_{-0.03})$,
$|R_{\psip}(0)|^2 = 0.93^{+0.17}_{-0.14}~\mathrm{GeV}^3$,
$|R_{\jpsi}(0)|^2 = 1.36^{+0.21}_{-0.18}~\mathrm{GeV}^3$,
$|R^\prime_{\chi_c}(0)|^2 = 0.08^{+0.01}_{-0.01}~\mathrm{GeV}^5$, and
$\left\langle \oo_{0}^{\chi_{c0}}\left({}^3S_1^{[8]}\right)\right\rangle = \left(0.36^{+0.05}_{-0.05}\right)\times10^{-2}~\mathrm{GeV}^3$, {close to, e.g., the NLO fit value of Ma et al.~\cite{Ma:2010vd}}.
We note that our fit yields a remarkable relative
precision 
on all six parameters, despite the large scale uncertainties.
Another important observation is that  $\delta_s = 0.07^{+0.16}_{-0.16}$ and $\delta_s^\prime = 0.05^{+0.06}_{-0.06}$
are both small and compatible with zero. The data therefore show no preference for a deviation  from the 
central  cross sections or from the central decay widths. Selecting the production data from higher $\pT$, 50 or 60 GeV, modifies the six LDMEs by less than 3\% with a similarly good $\chi_\text{data}^2$.

{We point out} that, by fitting cross-section ratios and polar
anisotropies---which are another type of cross-section ratio---along with three decay widths, the fit determines four LDMEs which act as global normalizations and two which drive the magnitude of the relativistic corrections. This means that the normalizations of the cross sections are genuine predictions of our fit, which we now confront with data.

\paragraph{Predictions using our LDMEs.}
Figure~\ref{fig:prediction_results} compares the absolute $\jpsi$ and
$\psip$ $\pT$-differential cross sections
to the ATLAS and CMS measurements.
These spectra are not included in the fit, but are predicted  using the optimal  values of the six LDMEs and our LP NLO+NLL calculations up to $\oo(v^2)$. The theoretical uncertainty bands are obtained by
propagating the parameter covariance matrix to the predicted cross sections. The excellent agreement
with the data provides an important consistency check of the extracted parameters, complementing
the constraints from production ratios and decay widths used in the fit.

\begin{figure}[!h]
    \centering
    \includegraphics[width=0.95\linewidth]{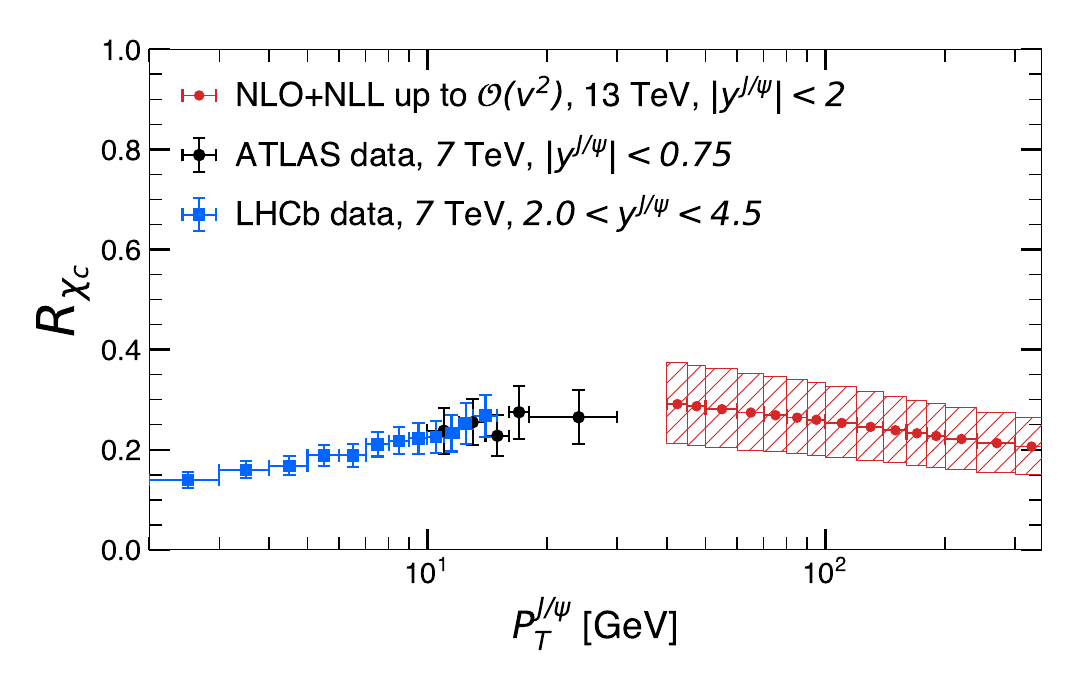}
    \caption{$\chic$ FD fraction  to  $\jpsi$ (red hatched) obtained with our six fit LDMEs, compared to the
    LHCb~\cite{LHCb:2012af} and ATLAS data~\cite{ATLAS:2014ala}.}
    \label{fig:chic-FD}
\end{figure}

With our LDMEs, one can also predict the {prompt} $\chi_c$ cross sections at high $\pT$ and the FD fraction into $J/\psi$, $R_{\chi_c}$. Figure~\ref{fig:chic-FD} compares our results with measurements by LHCb and ATLAS 
up to $\pT^{J/\psi}=25$~GeV. Our fit predicts that $R_{\chi_{c}}$ decreases with $\pT$, owing to the slower evolution of charm-fragmentation contribution to $\psi$ production. 

%
We now turn to another type of  predictions, considered for a long time
as a smoking gun of NRQCD, namely prompt $\jpsi$ polarization. 
Like for $\psip$,  this observable has been recently measured by CMS~\cite{CMS:2024igk}, up to $\pT=120$~GeV,
the main difference between both residing in the effect of FDs.
Following~\cite{Gong:2012ug}, we assume full polarization transfer in $\psi(2S)\to\jpsi X$, i.e.,
$\lambda^{\rm prompt}_{\theta,\psip}=\lambda^{\rm FD}_{\theta,\psip}$. This allows us to
account for the $\psip$ FD using the results shown in Fig.~\ref{fig:fit_results} (c).

In order to obtain the $\chic$ FD effect on $\lambda_{\theta, \jpsi}^{\rm prompt}$, we use
the CMS measurements~\cite{CMS:2019jas} of the relative $\chict / \chico$ FD polarization,
which give the following pairs for 
$(\lambda^{\rm FD}_{\theta, \chico}, \lambda^{\rm FD}_{\theta, \chict}) \in \{(0.0,-0.6),\, (1.0,0.3),\, (0.5,-0.6),\, (1.0,-0.3)\}$ as $68\%$ confidence boundary  of the corresponding 2D domain. %
\begin{figure}[!h]
    \centering
    \includegraphics[width=0.95\linewidth]{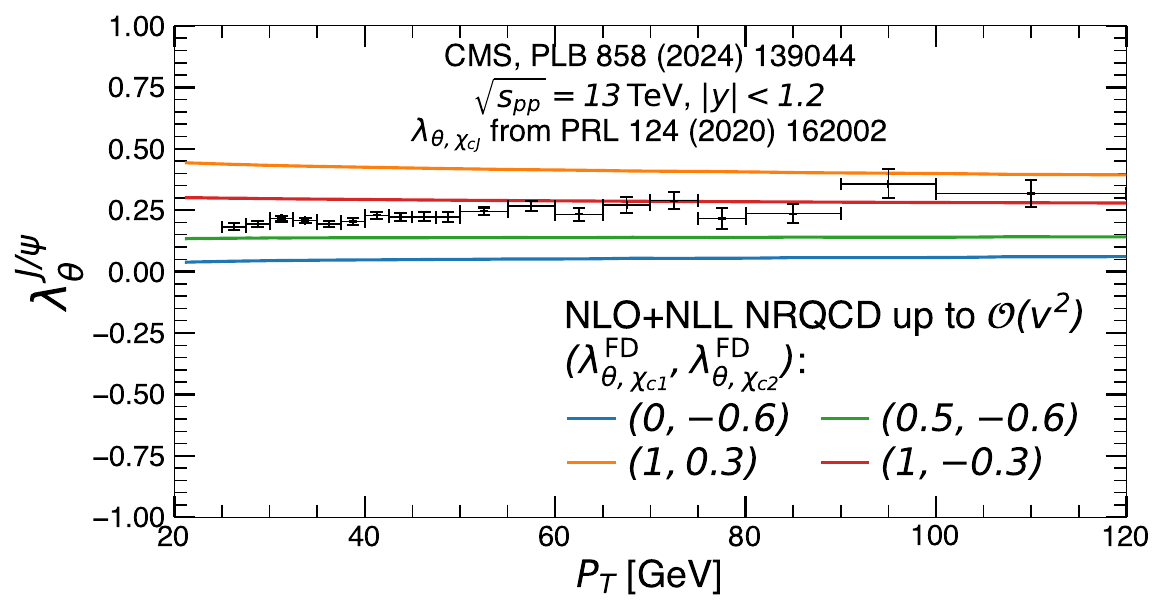}
    \caption{Polar anisotropy  $\lambda_\theta$ in the helicity frame for the prompt
    $\jpsi$ production measured by CMS at $\sqrt{s} = 13~\mathrm{TeV}$, $|y| < 1.2$, compared
    to our NLO+NLL predictions up to $\oo (v^2)$ using our six LDMEs.}
    \label{fig:jpsi_polarization}
\end{figure}

Figure~\ref{fig:jpsi_polarization} shows the resulting $\lambda_{\theta, \jpsi}^{\rm prompt}$
obtained at NLO+NLL using our LDMEs with the four
$\left(\lambda^{\rm FD}_{\theta,\chico}, \lambda^{\rm FD}_{\theta,\chict}\right)$
FD scenarios. Although the plot shows four colored bands, they appear as lines since uncertainties are extremely small.
We conclude that our NRQCD fit up to $\oo(v^2)$ also allows for an excellent description of the prompt $\jpsi$ polarization.

\paragraph{Conclusions.}
{We have performed a global determination of the charmonium average velocities squared
$\langle v^2_{\psi}\rangle$, of 
$|R_{\psi}(0)|^2$ and $|R^\prime_{\chi_c} (0)|^2$, as well as the CO
$\left\langle \oo_0^{\chi_{c}}\left({ }^3 S_1^{[8]}\right)\right\rangle$ LDME, by
combining LP NLO+NLL NRQCD at $\oo(v^2)$  with measured leptonic and radiative decay widths and two ratios of cross sections measured at large $\pT$ at the LHC. The first is the prompt $\psip/\jpsi$ yield ratio from ATLAS and CMS and the second is the polar anisotropy from CMS. }

{The best-fit values, $(\langle v^2_{\psip}\rangle, \langle v^2_{\jpsi}\rangle) = (0.38, 0.22)$, are in line with
the {expected hierarchy between ground and excited states}. By accounting for correlations  between scale uncertainties using nuisance parameters, we determine $|R_{\psi}(0)|^2$, $|R^\prime_{\chi_c}(0)|^2$ and $\left\langle \oo_0^{\chi_{c J}}\left({ }^3 S_1^{[8]}\right)\right\rangle$ with
remarkably small uncertainties.
The corresponding {values of  both} nuisance parameters indicate that the
best-fit {configuration} is only a few percent away from the central theoretical predictions.}
{Finally,} our fit provides a natural explanation why both {$\jpsi$ and $\psip$ polar anisotropies measured by CMS } are similar {and} slightly transverse.

{
The {quality} of our fit and its ability to correctly predict $\jpsi$, $\psip$, and $\chic$ hadroproduction cross sections at high $P_T$, where LP factorization is particularly robust,  demonstrate that NRQCD  truncated at $\oo (v^2)$ is sufficient to give a consistent description of production, including polarization, and decay. Such an agreement also  implies that higher $\oo (v^{n>2})$ corrections  proportional to CO $\psi$ LDMEs should be  small. In turn, this means that NRQCD converges faster than if one assumes that $\oo (v^{3,4})$ CO contributions to $\psi$ production are dominant.}

{Our fit value $|R_{\jpsi}(0)|^2 = 1.36^{+0.21}_{-0.18}~\mathrm{GeV}^3$ is slightly larger than  {previously assumed}, {i.e.,} $1.01~\mathrm{GeV}^3$ in, e.g.,~\cite{Ma:2008gq,Zhang:2006ay}, 
$0.95~\mathrm{GeV}^3$ in, e.g.,~\cite{Gong:2008ce,Gong:2009kp,Chang:2009uj},
$0.92~\mathrm{GeV}^3$ in, e.g.,~\cite{Kramer:1995nb,Butenschoen:2009zy} or $0.81~\mathrm{GeV}^3$ in, e.g.,~\cite{Artoisenet:2009xh}. A natural question is whether this improves or worsens the agreement with other inclusive $J/\psi$ production  observables described by NRQCD, whereas, by construction, it agrees with the $\jpsi$ leptonic decay width.

For total $P_T$-integrated  hadroproduction and photoproduction yields, such a slight increase of $|R_{\jpsi}(0)|^2$ will in general improve the agreement with data. Existing computations at NLO~\cite{Brodsky:2009cf,Feng:2015cba,Kramer:1995nb,ColpaniSerri:2021bla} (or NLO + double logarithmic resummation~\cite{Lansberg:2023kzf}) {exhibit} central-scale values slightly below the data. An increase of $|R_{\jpsi}(0)|^2$ will also improve the agreement with data for the $\pT$ differential photo-~\cite{Kramer:1995nb,Artoisenet:2009xh,Butenschoen:2009zy,Chang:2009uj,Flore:2020jau}  and {electroproduction}~\cite{Sun:2017wxk} cross sections. We note that  the corresponding $\oo(v^2)$ corrections are known to be small~\cite{Jung:1993cd,Schuler:1994hy,Xu:2012am}, except at the endpoint (e.g., $z\to 1$ for photoproduction).

The  same observation holds for $J/\psi$ production in $e^+e^-$ annihilation, where both $\sigma(J/\psi+X_{\text{non-}c\bar c})$ and $\sigma(J/\psi+X_{c\bar c})$ are known up to $\oo(\alpha\alpha_s^3+\alpha\alpha_s^2v^2)$, and their central-scale values~\cite{Zhang:2006ay,Ma:2008gq,Gong:2009kp,Gong:2009ng,He:2009uf} using  $|R_{\jpsi}(0)|^2\sim 0.9 -1.2~\mathrm{GeV}^3$ agree with the Belle data~\cite{Belle:2002tfa,Belle:2004abn,Belle:2009bxr}.
 $|R_{\jpsi}(0)|^2 = 1.36^{+0.21}_{-0.18}~\mathrm{GeV}^3$ will thus  maintain agreement with $e^+e^-$ annihilation data. 

Overall, it appears that NRQCD up to $\oo (v^2)$ describes the vast majority of charmonium data, including  $\eta_c$~\cite{LHCb:2014oii,LHCb:2019zaj} and $J/\psi+J/\psi$ hadroproduction~\cite{CMS:2014cmt,ATLAS:2016ydt,LHCb:2023ybt} (see~\cite{Butenschoen:2014dra,Han:2014jya,Zhang:2014ybe,Lansberg:2014swa,Lansberg:2019fgm,Lansberg:2019adr} and references therein). The only region which still calls for further investigation is  mid-to-large $\pT$ {for single $\psi$ hadroproduction}, where more in-depth studies of next-to-leading power contributions are needed. At the same time, studies of the $P_T$-integrated yields should be promoted to next-to-leading logarithmic accuracy along the lines of~\cite{Nefedov:2024swu,Fucilla:2026ooq}.}

\begin{acknowledgments}
{\it Acknowledgments --- }
We are indebted to A.P.~Chen, Y.Q.~Ma, F.~Feng, Y. Jia, C.Y. Liu, J. Gao and D. Yang for their assistance in using their codes or results. We warmly thank M.~Fontannaz, M.~Nefedov and H.~Sazdjian  for their numerous inputs and comments, as well as  M. Cacciari,   C.~Flett, V. Kartvelishvili, C.F.~Qiao, L.~Maxia, H.S.~Shao, L.~Simon, P.~Taels, C.~Van Hulse and T.~Zakareishvili for discussions. 

{\it Funding --- }The research conducted in this publication was funded by the Irish Research Council under grant number GOIPG/2022/478.
This project has received funding from the European Union’s Horizon Europe research and innovation programme under the grant agreement No.~101292522 (HADRON-2030) in order to contribute to the EU Virtual Accesses NLOAccess and 3DPortal. 
This project has also received funding from the Agence Nationale de la Recherche (ANR) via the grants ANR-20-CE31-0015 (``PrecisOnium''), ANR-24-CE31-7061-01 (``3DLeaP"), and via the IDEX Paris-Saclay ``Investissements d'Avenir" (ANR-11-IDEX-0003-01) through the GLUODYNAMICS project funded by the ``P2IO LabEx (ANR-10-LABX-0038)" and through the Joint PhD Programme of Universit\'e Paris-Saclay (ADI). This work  was also partly supported by the French CNRS via the IN2P3 projects ``GLUE@NLO" and ``QCDFactorisation@NLO".
\end{acknowledgments}

\bibliographystyle{utphys}
\bibliography{bib}

\end{document}